\documentclass[review]{elsarticle}
\usepackage{lineno,hyperref,bm}
\modulolinenumbers[5]

\begin{document}

\begin{frontmatter}

\title{Devitalizing noise-driven instability of entangling logic in silicon devices with bias controls}

\author{Hoon Ryu\corref{cor}}
\author{Ji-Hoon Kang}
\cortext[cor]{Corresponding author (Email address: elec1020@kisti.re.kr)}
\address{Division of National Supercomputing,\\Korea Institute of Science and Technology Information, Daejeon 34141, Republic of Korea}

\begin{abstract}
The quality of quantum bits (qubits) in silicon is highly vulnerable to charge noise that is omni-present in semiconductor devices and is in principle hard to be suppressed. For a realistically sized quantum
dot system based on a silicon-germanium heterostructure whose confinement is manipulated with electrical biases imposed on top electrodes, we computationally explore the noise-robustness of 2-qubit
entangling operations with a focus on the controlled-X (CNOT) logic that is essential for designs of gate-based universal quantum logic circuits. With device simulations based on the physics of bulk semiconductors
augmented with electronic structure calculations, we not only quantify the degradation in fidelity of single-step CNOT operations with respect to the strength of charge noise, but also discuss a strategy of
device engineering that can significantly enhance noise-robustness of CNOT operations with almost no sacrifice of speed compared to the single-step case. Details of device designs and controls that this
work presents can establish a rare but practical guideline for potential efforts to secure silicon-based quantum processors using an electrode-driven quantum dot platform.
\end{abstract}

\begin{keyword}
Noise-robust entangling operations \sep Electrode-driven quantum dot structures \sep Silicon-based quantum computing \sep Computational nanoelectronics
\end{keyword}

\end{frontmatter}


\section{Introduction}
The spin of electrons in isotopically enriched silicon (Si) has been regarded as a promising mechanism for encoding quantum information due to its extremely long coherence time \cite{R01,R02,R03,R04}
that is highly advantageous for stable manipulations of quantum bits (qubits). In particular, a great amount of effort has been put in by researchers to physically realize universal logic gate devices with electron
spins in Si-based quantum dot (QD) structures \cite{R03,R04,R05,R06,R07,R08,R09,R10,R11} whose confinement is controlled with external electric and magnetic fields. The preciseness of corresponding
logic operations has been continuously improved so single qubit rotations can be now conducted with a fidelity larger than 99\% \cite{R03,R04,R05,R06,R07,R11}, and recently a 98\% fidelity is reported for
2-qubit SWAP operations \cite{R08}. Elaborated designs of gate devices that generate quantum entanglement \cite{R12,R13}, the most celebrated quantum resource being widely used in various applications
\cite{R14,R15,R16}, have been also reported but their accuracy so far is generally not as good as the non-entangling cases so the fidelity of 2-qubit Bell-states generated from double quantum dot (DQD)
platforms stays in 78\%-89\% \cite{R09,R10,R11}. With a rapid progress in pulsing technologies \cite{R06,R17}, the speed of DQD-based gating operations reached a sub-microsecond level, and it is shown
that a controlled-X (CNOT) operation, the most crucial entangling logic for universal quantum computing, can be conducted in less than 200 nanoseconds (nsec) with a single microwave pulse \cite{R10}.

In general, the quality of spin qubits in solid-based platforms highly depends on material-inherent noises \cite{R18,R19,R20,R21,R22} that are mainly due to the fluctuation in local electric and magnetic fields
around qubits. In the Si-based case, noises of magnetic fields (spin noises) can be suppressed with purification of $^{28}$Si  crystals, and latest works have shown 12-inch $^{28}$Si wafers that contain 100ppm
or less of spin-carrying $^{29}$Si atoms \cite{R23,R24}. Suppressing noises of electric fields (charge noises), however, is more difficult than the case of spin noises since its origins have not been fully understood
yet. Accordingly, state-of-the-art ideas have been proposed to increase the robustness of spin qubits to existing charge noises in Si devices such as, for example, placing qubits far away from surface oxides
\cite{R22} that can serve as a source of low-frequency charge noises \cite{R19}, increasing spin resonance frequencies \cite{R07}, and biasing DQDs symmetrically to reduce the sensitivity of qubit interactions
to charge noises \cite{R10,R25}. In spite of the non-trivial contribution driven with these ideas, the up-to-date fidelity of entangling operations in Si devices is not yet in a level where the accuracy in computations
can be generally guaranteed, and the motivation for sound studies on technical strategies that can enhance the fidelity of entangling operations under charge noises, therefore, should be huge.

In this work, we elaborately examine the engineering-driven possibilities for devitalizing negative effects that charge noises have against entangling operations implemented with Si QD devices, where the focal
point of engineering is the real-time pattern of control signals that has been rarely discussed in detail by the strategies proposed in previous studies \cite{R07,R22,R25}. For this purpose, we computationally
explore Si DQD structures with our in-house simulation code package that can describe device operations in a full-scale from initializations and time-dependent behaviors of electron spin qubits. As a baseline
for discussion, we first model the fast CNOT operation that is driven with a single-step pulse in the recently reported DQD platform \cite{R10}, and quantify fluctuations in fidelity under charge noises by incorporating
random noisy potential profiles to device simulations. Then, as an alternative way, we implement a CNOT operation with a multi-step control that does not employ AC microwave pulses for generation of entanglement.
In spite of the loss in fidelity that happens during the real-time transition of control signals, we find a general pattern that the resulting CNOT operation has remarkably increased robustness to charge noises whilst
its operating speed can be maintained in a same order compared to the case of a single-step control. Additional in-depth discussion is presented via rigorous modeling to study the optimal control of multi-step
CNOT operations in realistic conditions with a trade-off between the speed and the noise-robustness of operations. Being carried as an extension of our preliminary study that focused on the noise-free addressing
of individual qubits \cite{R26}, this work can make a meaningful contribution for Si-based designs of entangling logic blocks that are essential for development of programmable quantum processors.

\section{Methods}
\begin{figure}[t]
\centering
\includegraphics[width=\columnwidth]{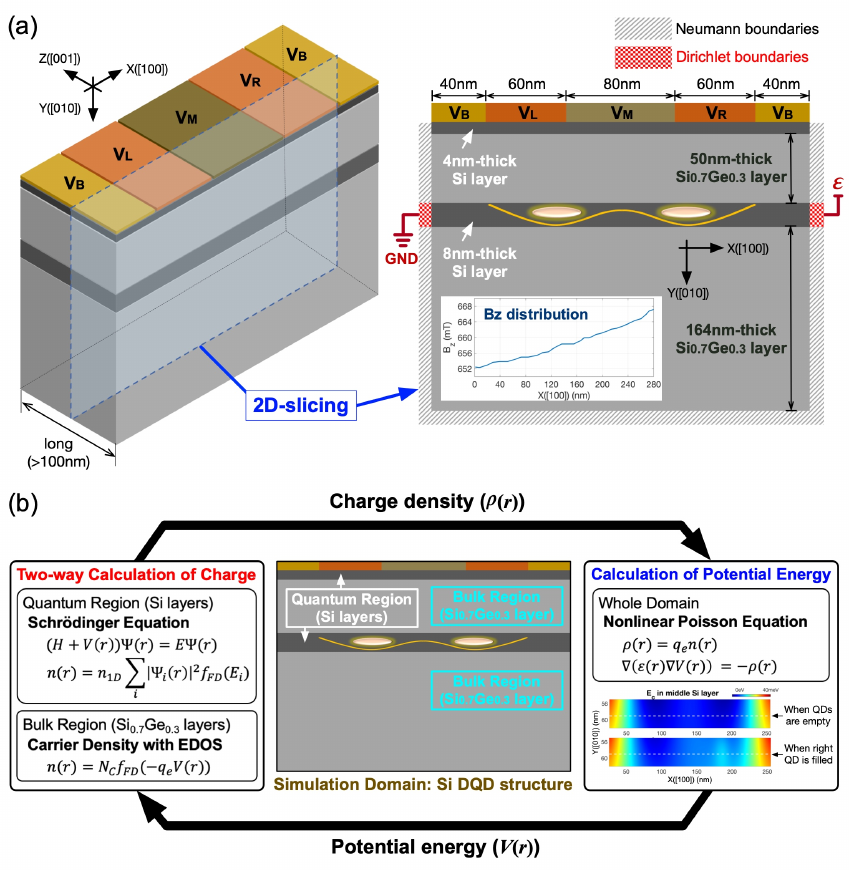}
\caption{\textbf{Target structure and multi-scale scheme of device modeling.} (a) A 3D view of the silicon (Si) double quantum dot (DQD) structure that resembles the physical one reported by Zajac $et$ $al.$ (Ref. [10]). Here the
	lateral confinement (along the [100] direction) is controlled with DC biases imposed on the top electrodes, while the vertical one (along the [010] direction) is naturally formed due to the band offset among silicon-germanium
	(SiGe) and Si layers. Since the structure is very long along the [001] direction, we use its 2D slice for device simulations assuming it is infinitely long along that direction (a lateral distribution of the static magnetic field $B_Z$,
	generated from a horseshoe-shaped cobalt micromagnet, is shown in the inset). (b) The self-consistent loop of device simulations used to model spatial distributions of charge and potential. Here the charge distribution at a
	given potential distribution is obtained in two ways; the electronic structure simulation based on a parabolic effective mass model is used to get the density in Si layers where most of electrons reside, while the region of SiGe
	layers is treated with the physics of bulk semiconductors.
	}
\label{fig:met1}
\end{figure}

Figure \ref{fig:met1}(a) shows the DQD structure that is adopted as a target of modeling in this work. Mimicking the reported physical system \cite{R10}, the target platform is based on a heterostructure that consists
of 2 Si and 2 silicon-germanium (SiGe) layers where the fraction of Ge in SiGe layers is 30\%. Due to the Si$/$SiGe band offset, the structure has a natural quantum well along the vertical ([010]) direction and
electrons can be confined in the 8 nanometer(nm)-thick Si layer. The lateral ([100]) confinement in the 8nm-thick Si layer is controlled with DC biases imposed on top Ti$/$Au electrodes (2 barrier gate biases
($V_B$), 1 left$/$middle$/$right gate bias ($V_L$$/$$V_M$$/$$V_R$)), so the system can have up to 2 potential valleys. As the DQD system is quite long ($>$100nm) along the [001] direction, we use its 2D-slice
as a simulation domain assuming the structure is infinitely long along that direction. The top electrodes are considered in device simulations by imposing a Dirichlet boundary condition on a 2D Poisson equation
with applied biases and Schottky barrier heights ($\Phi_B$) that are calculated using the work-function reported for Ti/Au metal layers. \cite{R27} The source and drain electron reservoirs, which are secured with
2D electron gas (2DEGs) in reality, are also described with Dirichlet boundaries (the two red boundaries in Figure \ref{fig:met1}(a)) where we set $\Phi_B$ to zero assuming that 2DEGs are formed well and are
therefore perfectly ohmic. For simulations, we grounded the source and imposed an extremely small bias ($\varepsilon$ $\simeq$ 0.1mV) on the drain, and a low temperature of 1.5K is assumed.

\begin{figure}[t]
\centering
\includegraphics[width=\columnwidth]{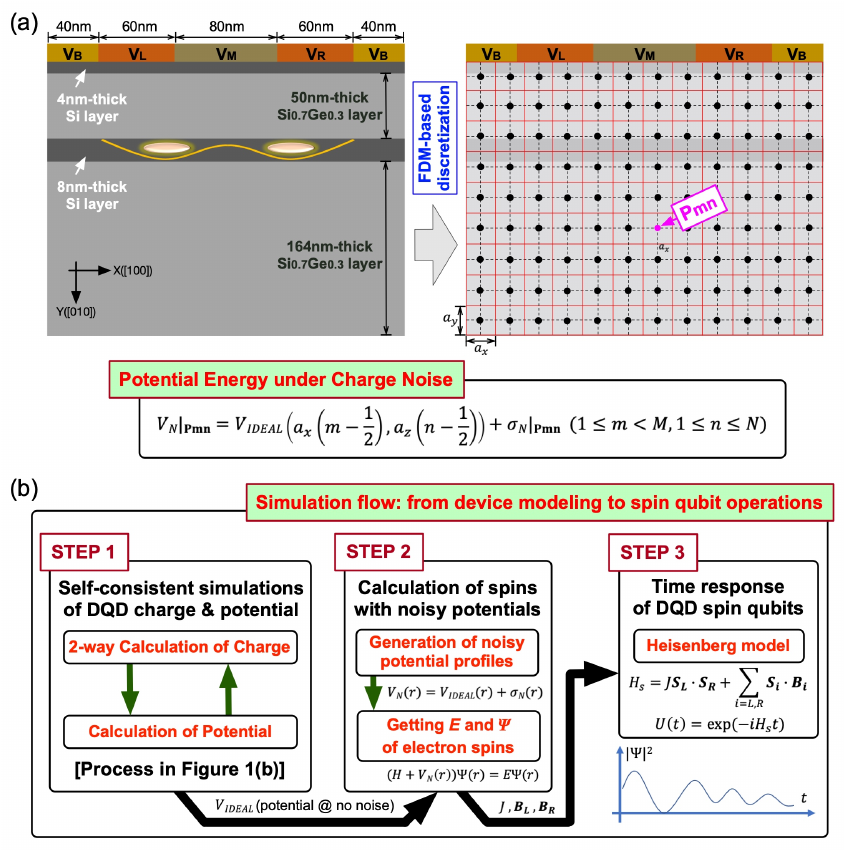}
\caption{\textbf{Incorporation of charge noise and steps for full-stack simulations covering from device modeling to qubit operations.} (a) The effect of charge noise is incorporated into device simulations by disturbing the spatial
	distribution of potential energy under no noise ($V_{IDEAL}$) with a random value that is generated per real-space grid in a silicon (Si) double quantum dot (DQD) structure under a zero-mean gaussian distribution of a
	standard deviation up to 5$\mu$eV. (b) Starting with device modeling, our full-stack simulation can eventually predict 2-qubit operations of a Si DQD system with the following three steps: (1) device simulations that give
	bias-dependent energetic positions and wavefunctions of electron spin states under no charge noise, (2) disturbing the noise-free potential profile with charge noise, and (3) solving a time-dependent Schr\"odinger equation
	for the Heisenberg Hamiltonian of two neighbor spins that is constructed with results of device simulations.
	}
\label{fig:met2}	
\end{figure}

The spatial distribution of potential energy and electron density in the DQD system, which is the outcome of device simulations, is determined with a self-consistent process described in Figure \ref{fig:met1}(b).
While the potential profile is calculated with a normal Poisson solver, the charge profile is evaluated in two ways with regional dependence so the region of thin Si layers (labeled as \textbf{Quantum Region}),
which has most of electrons and must be solved quantum mechanically, is treated with electronic structure simulations coupled to a parabolic effective mass model \cite{R28}, and the region of SiGe layers
(labeled as \textbf{Bulk Region}) is solved with the physics of bulk semi-conductors. For precise modeling of spin states, the electronic structure is calculated with a lateral distribution of the static magnetic
field along the [001] direction ($B_Z$) that is reported by Neumann $et$ $al.$ \cite{R29} with simulations of the horseshoe-shaped micromagnet employed in the real experiments \cite{R10,R30,R31} (see
the inset of Figure \ref{fig:met1}(a)). Once the potential distribution at a certain set of biases is determined, we disturb this ``clean'' solution with a noisy potential profile, which is obtained with values that are
randomly generated per each real-space grid of the simulation domain as described in Figure \ref{fig:met2}(a). All the random values here are generated under a zero-mean gaussian distribution, and its standard
deviation $\sigma$, which represents the strength of charge noises, is considered up to 5$\mu$eV that is normal in Si-based devices these days \cite{R32,R33,R34,R35}. Once the ground states of two QDs
are known from device simulations, we can construct the Heisenberg 2-spin Hamiltonian with their Zeeman-spitting energies and exchange interaction \cite{R36}, and 2-qubit time responses of the DQD system
can be then calculated as described in Figure \ref{fig:met2}(b) that shows the scheme of our full-stack modeling.

\section{Results and Discussion}
\begin{figure}[t]
\centering
\includegraphics[width=\columnwidth]{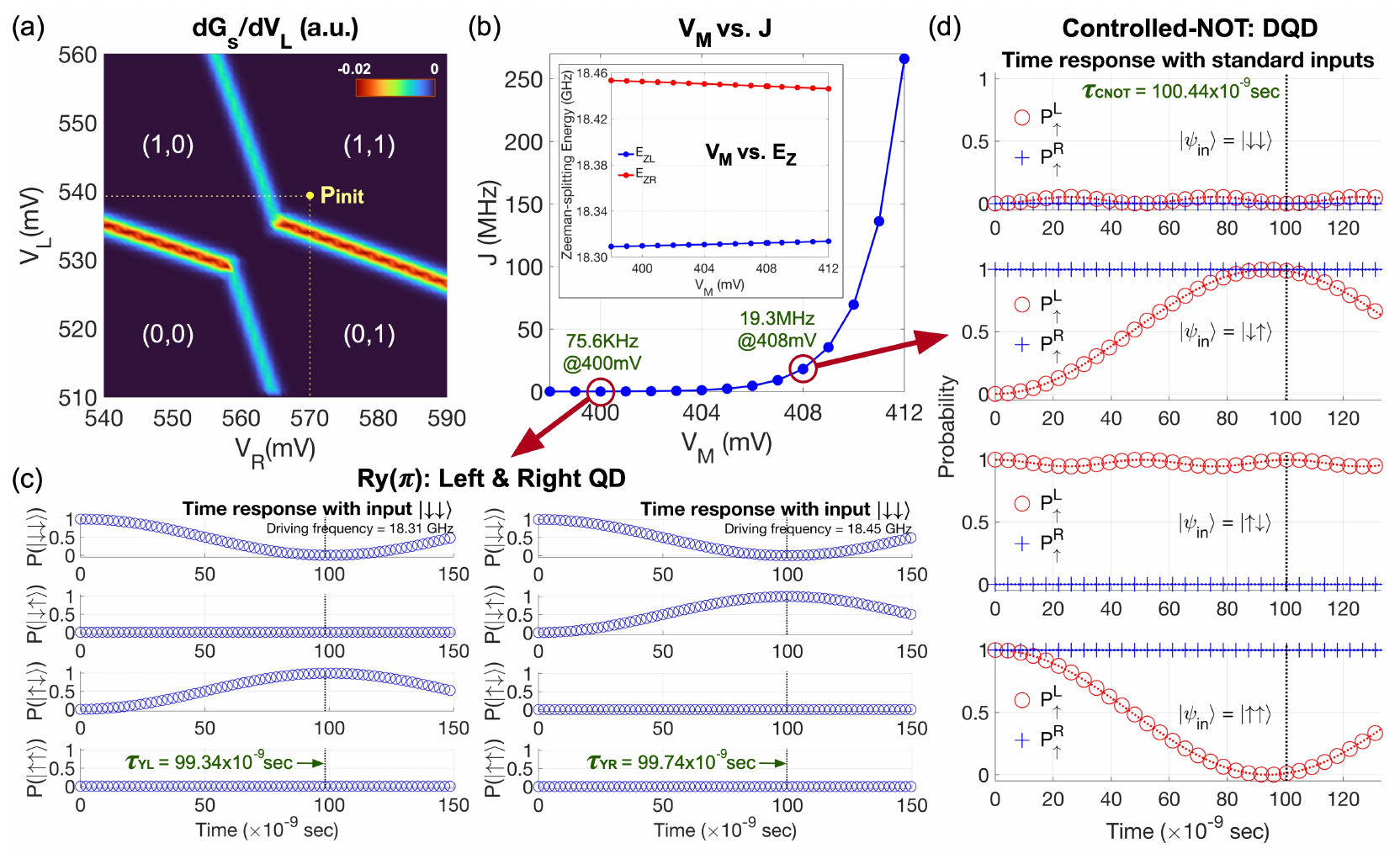}
\caption{\textbf{Device initialization, 1-qubit rotation and single-step CNOT operation.} (a) A charge stability diagram of the double quantum dot (DQD) system that shows electron-filling in each QD as a function of the left \& right
	gate bias ($V_L$ \& $V_R$). The middle gate bias ($V_M$) is set to 400mV. With controls of gate biases, we can fill a single electron in each QD, initializing the device to a $|$$\downarrow\downarrow$$\rangle$ state. (b)
	Exchange ($J$) and Zeeman-splitting energy ($E_{ZL}$, $E_{ZR}$) of two QDs shown as a function of $V_M$ when $V_L$ and $V_R$ are 540mV and 570mV, respectively. Increasing $V_M$ lowers the potential barrier
	between two QDs and enhances the interaction between electrons that occupy the down-spin ground state of each QD. (c) Time responses of the DQD system at $V_M$ = 400mV. The $R_Y(\pi)$ operation for the left
	and right qubit is completed in 99.34 and 99.74 nanoseconds (nsec), respectively. When the interaction is weak, we can address each qubit independently by setting the frequency of an AC microwave pulse equal to the
	ground state Zeeman-splitting energy of each QD, which is 18.31GHz (left) and 18.45GHz (right) in our case. (d) Time responses simulated at $V_M$ = 408mV that achieve a single-step completion of the controlled-NOT
	operation in 100.4nsec. A 8mV increase of $V_M$ dramatically enhances the interaction of QDs so $J$ at $V_M$ = 408mV is $\sim$250 times larger than the case of $V_M$ = 400mV. Once QDs strongly interact, the Rabi
	frequency of the qubit in one QD depends on the state of the qubit in the other QD, generating 2-qubit entanglement.
	}
\label{fig:res1}	
\end{figure}

In any physical platforms, the first step for gating operations is to initialize qubits so the system can be prepared for upcoming control pulses. In the target DQD platform where a qubit state 0 and 1 are encoded
to the down-spin ($|$$\downarrow$$\rangle$) and the up-spin ($|$$\uparrow$$\rangle$) ground state of a QD, respectively, initialization is done by manipulating biases imposed on top electrodes such that the
$|$$\downarrow$$\rangle$ state in each QD is occupied with an electron. To quantify the range of biases that can place the target system in the ($|$$\downarrow$$\rangle_L$, $|$$\downarrow$$\rangle_R$)
state (= $|$$\downarrow$$\downarrow$$\rangle$) where the subscription $L$ and $R$ represent for the left and right QD, respectively, we model the charge stability with device simulations, and present the
result in Figure \ref{fig:res1}(a) as a function of $V_L$ and $V_R$ at $V_M$ = 400mV, where $V_B$ is fixed to 200mV. The stability diagram is split to 4 regimes, and each one is identified with two numbers
that represent the electron population of each QD. With increasing $V_{L(R)}$, the ground state of the left(right) QD shifts down in energy and is occupied with an electron when the state touches the Fermi-level
of the source electron reservoir. Establishing a strong connection to data measured for the physical DQD system \cite{R10}, our result reveals that ($V_L$, $V_R$) = (540mV, 570mV) (the yellow point labeled
as \textbf{Pinit}) can be an initialization point that is beneficial for noise-robust qubit interactions since two QDs can be symmetrically biased \cite{R25}.

Representing the strength of inter-QD qubit interaction, the exchange energy ($J$) between two ground states serves as a source of 2-qubit entanglement in the DQD platform and can be controlled with the
middle gate bias that affects the potential barrier between two QDs. In our simulations, $J$ becomes 75.6KHz at $V_M$ = 400mV (at \textbf{Pinit}) and, as shown in Figure \ref{fig:res1}(b), sharply reaches
19.3MHz when $V_M$ is increased by 8mV. Changes in $V_M$ also affect Zeeman-splitting energies of the left ($E_{ZL}$) and the right ground state ($E_{ZR}$) that determine the resonance frequency of
each spin qubit, but their dependence on $V_M$ is not quite noticeable such that ($E_{ZL}$, $E_{ZR}$) is (18.309GHz, 18.453GHz) at $V_M$ = 400mV and changes to (18.312GHz, 18.448GHz) when $V_M$
is 408mV. Due to the laterally inhomogeneous $B_Z$ (the inset of Figure \ref{fig:met1}(a)), $E_{ZL}$ and $E_{ZR}$ are distinguishable and qubits can be addressed independently if their interaction is weak,
and one of such cases is shown in Figure \ref{fig:res1}(c), where we simulated 2-qubit responses at $V_M$ = 400mV with a [010]-oriented time-varying magnetic field $B_Y$${(t)}$ = $B_o$${cos(\omega_D t + \theta)}$
that is generated from a microwave pulse and is incorporated in modeling as elements of the Heisenberg Hamiltonian. In particular, the two subfigures here show that a $R_Y(\pi)$ operation (1-qubit rotation
by $\pi$ radian around the Y-axis) can be selectively implemented with each qubit by setting $\omega_D$ to $E_{ZL}$ or $E_{ZR}$, and gating is completed in 99.34 \& 99.47nsec (left \& right) when $B_o$
= 5.0MHz and $\theta$ = 0. If the interaction is not weak enough to ignore, the resonance frequency of one qubit starts to depend on the spin state ($|$$\downarrow$$\rangle$ or $|$$\uparrow$$\rangle$) of
the other qubit, and a CNOT operation can be then realized with a single control pulse \cite{R10,R36}. To mimic this 1-step implementation with modeling, we simulate the DQD structure at $V_M$ = 408mV
with $B_Y$${(t)}$ of $\omega_D$ (= 1.832GHz), $B_o$ (= 4.977MHz) and $\theta$ (= 1.5$\pi$ radian) that are determined with the analytical solution driven by Russ $et$ $al.$ \cite{R36}. Simulated 2-qubit
responses in Figure \ref{fig:res1}(d) clearly show the CNOT gating is completed in 100.4nsec, being fairly connected to the experiment \cite{R10}.

\begin{figure}[!t]
\centering
\includegraphics[width=\columnwidth]{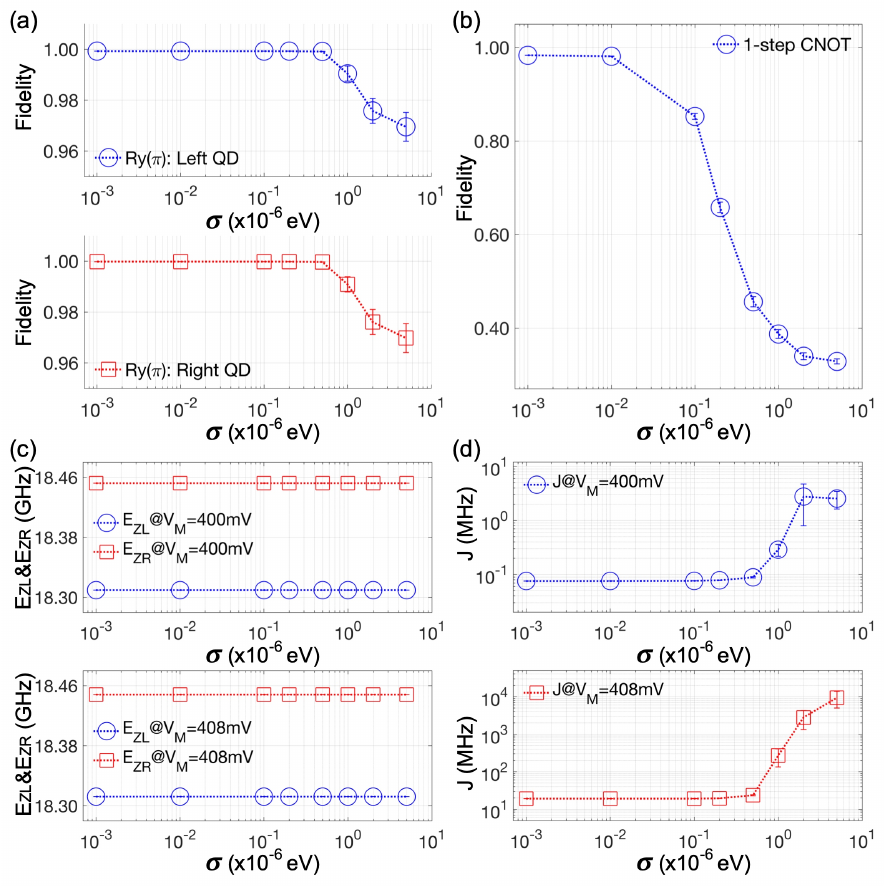}
\caption{\textbf{Noise-driven behaviors of 1-qubit rotation and single-step CNOT operation.} (a) The fidelity of a $R_Y(\pi)$ operation conducted with the left and the right quantum dot (QD) are presented as a function of the
	magnitude of charge noise ($\sigma$: standard deviation of noisy potential values that are randomly generated per grid of the modeling domain). The fidelity obtained with 1,000 samples is 99.93$\pm$10$^{-6}$\% (left
	QD) and 99.98$\pm$10$^{-6}$\% (right QD) when $\sigma$ = 10$^{-3}\mu$eV, and becomes 96.95$\pm$0.5664\% and 96.98$\pm$0.5687\% when $\sigma$ = 5$\mu$eV. (b) The single-step controlled-NOT operation
	turns out to be much more vulnerable to charge noise than the case of 1-qubit rotations so the fidelity becomes 98.34$\pm$0.003\% and 32.84$\pm$0.5361\% when $\sigma$ is 10$^{-3}\mu$eV and 5$\mu$eV, respectively.
	(c) Noise-driven fluctuation in Zeeman-splitting energy ($E_{ZL}$, $E_{ZR}$) and (d) exchange interaction ($J$) is shown in the regime of weak ($V_M$ = 400mV) and strong interaction ($V_M$ = 408mV), revealing that
	the degradation in fidelity, particularly in the case of entangling operation, is due to the sensitivity of $J$ to charge noise.
	}
\label{fig:res2}	
\end{figure}

In a noise-free condition, modeling results show that the $R_Y(\pi)$ operation is conducted for the left and the right spin with a fidelity of 99.93\% and 99.98\%, respectively, and the 1-step CNOT operation has
a fidelity of 98.34\%. To investigate how they are affected by charge noises, we simulate the system with the conditions described in the previous paragraph but disturb the clean potential profiles with random
noisy values that are generated under a zero-mean gaussian distribution of a standard deviation $\sigma$. Figure \ref{fig:res2}(a) and \ref{fig:res2}(b) show the fidelity of the $R_Y(\pi)$ and the 1-step CNOT
operation as a function of $\sigma$, respectively, where each case is modeled by conducting 1,000 simulations per a single value of $\sigma$ that is varied from 10$^{-3}$ to 5$\mu$eV. Results clearly indicate
that both operations continue to lose accuracy as the DQD system experiences more severe noises, but their patterns of the noise-driven degradation in fidelity are different. In the case of $R_Y(\pi)$ gating,
the fidelity turns out to be 99.93$\pm$10$^{-6}$\% (left) and 99.98$\pm$10$^{-6}$\% (right) at $\sigma$ = $10^{-3}$$\mu$eV, and starts to decrease noticeably when $\sigma$ reaches 1$\mu$eV or larger
such that it drops to 96.95$\pm$0.5664\% (left) and 96.97$\pm$0.5687\% (right) at $\sigma$ = 5$\mu$eV, Similarly to the $R_Y(\pi)$ case, the 1-step CNOT operation has a nice fidelity (98.34$\pm$0.003\%)
when $\sigma$ is $10^{-3}$$\mu$eV. Its robustness to noises however is much worse than what $R_Y(\pi)$ shows, and the average fidelity plummets more than 60\% (32.84$\pm$0.5361\%) when $\sigma$
is 5$\mu$eV.

In the extreme case where the left and right qubit never interact ($i.e.$, $J$ = 0), the Heisenberg 2-spin Hamiltonian can be completely described with Zeeman-splitting energies of spin states and external
magnetic fields, and so are 2-qubit responses of the DQD system. Accordingly, in the regime of a weak interactions that can be represented with the case of $V_M$ = 400mV, the quality of single qubit addressing
under charge noises should be determined by how $E_{ZL}$ and $E_{ZR}$ behave. The origin of noise-robust $R_Y(\pi)$ rotations (Figure \ref{fig:res2}(a)) can be therefore clarified with simulation results
presented in the upper subfigure of Figure \ref{fig:res2}(c), which indicate that the noise-driven fluctuation in two Zeeman-splitting energies at $V_M$ = 400mV becomes smaller than 100Hz (10$^{-5}$\% of
their clean values) regardless of $\sigma$. If the 2-qubit interaction is not ignorable as it is when $V_M$ = 408mV, $J$ also starts to affect the noise-robustness of gating. As Figure \ref{fig:res2}(d) shows, the
noise-driven fluctuation in $J$ is generally much stronger than the $E_{ZL}$ \& $E_{ZR}$ case, and, particularly at $V_M$ = 408mV, it acts as the major factor that determines the noise-robustness of 2-qubit
states because our results reveal that the fluctuation in $E_{ZL}$ and $E_{ZR}$ is still negligible as shown in the lower subfigure of Figure \ref{fig:res2}(c). In consequence, we can say that the huge reduction
in fidelity observed in noisy 1-step CNOT operations (Figure \ref{fig:res2}(b)) is mainly due to the noise-driven instability of $J$.

\begin{figure}[t]
\centering
\includegraphics[width=\columnwidth]{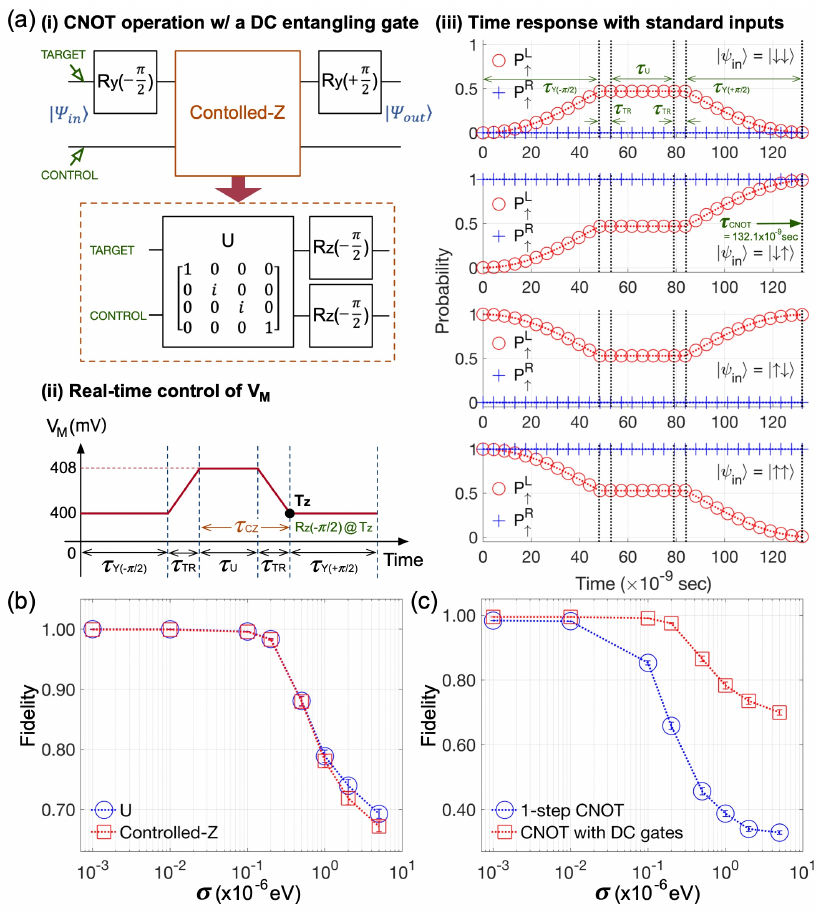}
\caption{\textbf{Multi-step CNOT operation with DC entangling logic.} (a) (i) The CNOT operation can be achieved with three steps, where the second one (a controlled-Z (CZ) gate) consists of a 2-qubit unitary $U$ and two 1-qubit
	rotation blocks ($R_Z(-\pi/2)$) that can be implemented with only DC biases in double quantum dot (DQD) platforms. Here the entanglement is generated by $U$. (ii) A real-time control of the middle gate bias ($V_M$) for the
	three-step CNOT operation when the left and right gate bias are 540mV and 570mV, respectively. A bias-transition time ($\tau_{TR}$) of 5 nanoseconds (nsec) is assumed. (iii) Resulting 2-qubit time responses show that the
	CNOT operation is completed at 132.1nsec. (b) The fidelity of a $U$ and a CZ block are shown as a function of the magnitude of charge noise, indicating that 1-qubit Z-rotations do not quite affect the preciseness of the overall
	CZ operation. (c) The multi-step CNOT operation is much more robust to charge noise than the single-step case, mainly due to the noise-robustness of the DC entangling block $U$.
	}
\label{fig:res3}	
\end{figure}

Given that the material-inherent charge noise itself would not be easy to be eliminated or hugely suppressed, the next action for implementation of reliable CNOT operations may be to seek for engineering
approaches that can make the gate more robust to ``existing'' noises. For this purpose, here we computationally explore one idea whose main focus is to control qubit interactions such that the ``noise-sensitive''
interval in time responses can be reduced as much as possible. In Figure \ref{fig:res3}(a)-(i), we show a simple 2-qubit circuit which also conducts a CNOT operation and will be used as a testbed of the
noise-robustness. Here, the desired gating can be implemented with a time-sequential conduction of a $R_Y(-\pi/2)$, a controlled-Z (CZ), and a $R_Y(\pi/2)$ operation where $R_Y$ rotations are applied
to the upper (target) qubit. Taking the right QD spin as a control qubit, we can implement the two $R_Y$ gates in the DQD platform at $V_M$ = 400mV by setting $B_Y$${(t)}$ similarly to the $R_Y(\pi)$
case except that $\theta$ is $\pi$ (instead of 0) when the rotation angle is negative. The CZ gate in the second step serves as an entangling block and can be obtained without time-varying microwave pulses.
Technically, a CZ gate can be further decomposed into 2 steps as illustrated in the bottom subfigure of Figure \ref{fig:res3}(a)-(i). Here, the 2-qubit controlled-phase gate $U$ is $device$-$native$ \cite{R36},
which means the unitary can be completely described with only DQD-native spin parameters ($i.e.$, Zeeman-splitting energies and exchange interaction). The Z-rotation ($R_Z$) is also device-native but
must be carried in the regime of a weak interaction ($e.g.$, $V_M$ = 400mV in our case). In real experiments, the $R_Z$ is conducted by changing the reference phase for individual spins instead of directly
rotating them, which can be done conveniently with software at negligible cost in speed and accuracy \cite{R07,R09,R10}. In Figure \ref{fig:res3}(a)-(ii), we show the real-time pattern of $V_M$ that drives
this multi-step CNOT gate, where $\tau_{Y}$'s and $\tau_{U}$ on the X-axis are the gating time of $R_Y(\pm\pi/2)$ and $U$, respectively. We assume that the $R_Z$ gating is performed instantaneously
(at the time point labeled as $\bf{T_Z}$), adopting a bias-transition time ($\tau_{TR}$) of 5nsec for simulations similarly to the experiment \cite{R10}. The resulting responses in Figure \ref{fig:res3}(a)-(iii)
reveal that the entire process takes 132.1nsec, where $\tau_{Y}(\pm\pi/2)$ and $\tau_U$ become 48.1nsec and 25.9nsec, respectively.

The focal characteristic of the above-mentioned multi-step CNOT operation is that the 2-qubit entanglement is solely generated by the CZ block, and eventually by the controlled-phase unitary $U$, as all
the remaining logics ($R_Y$'s and $R_Z$'s) handle 1-qubit addressing in the regime of a weak interaction. As the sensitivity of $E_{ZL}$ and $E_{ZR}$ to charge noises is not quite noticeable (Figure \ref{fig:res2}(c)),
the fairly nice noise-robustness of $R_Y$ shown in Figure \ref{fig:res2}(a) also becomes valid for 1-qubit rotations about arbitrary axes. We can thus expect that the noise-driven fidelity of the CZ operation
may strongly depend on that of $U$, and this can be confirmed with Figure \ref{fig:res3}(b) that shows the simulated pattern in fidelity of CZ and $U$ gating. Due to the negligible role of $R_Y$ blocks, the
overall fidelity of the multi-step CNOT logic, shown with a red dotted line of square marks in Figure \ref{fig:res3}(c), also closely follows the fidelity of $U$. When $V_M$ is 408mV, the multi-step CNOT logic
generates 2-qubit entanglement in $\sim$4x less time (25.9nsec) than the 1-step gating (100.4nsec). This ``reduced time-period of a strong interaction'' can contribute to making the operation more robust
to charge noises, so the simulated fidelity of the multi-step operation at $\sigma$ = 5$\mu$eV becomes 69.81$\pm$0.8208\% while the 1-step CNOT gate shows 32.84$\pm$0.5361\% in the same conduction.
Our result in Figure \ref{fig:res3}(c) also confirms the core message remains effective in the entire range of $\sigma$ that is considered for simulations.

\begin{figure}[ht]
\centering
\includegraphics[width=\textwidth]{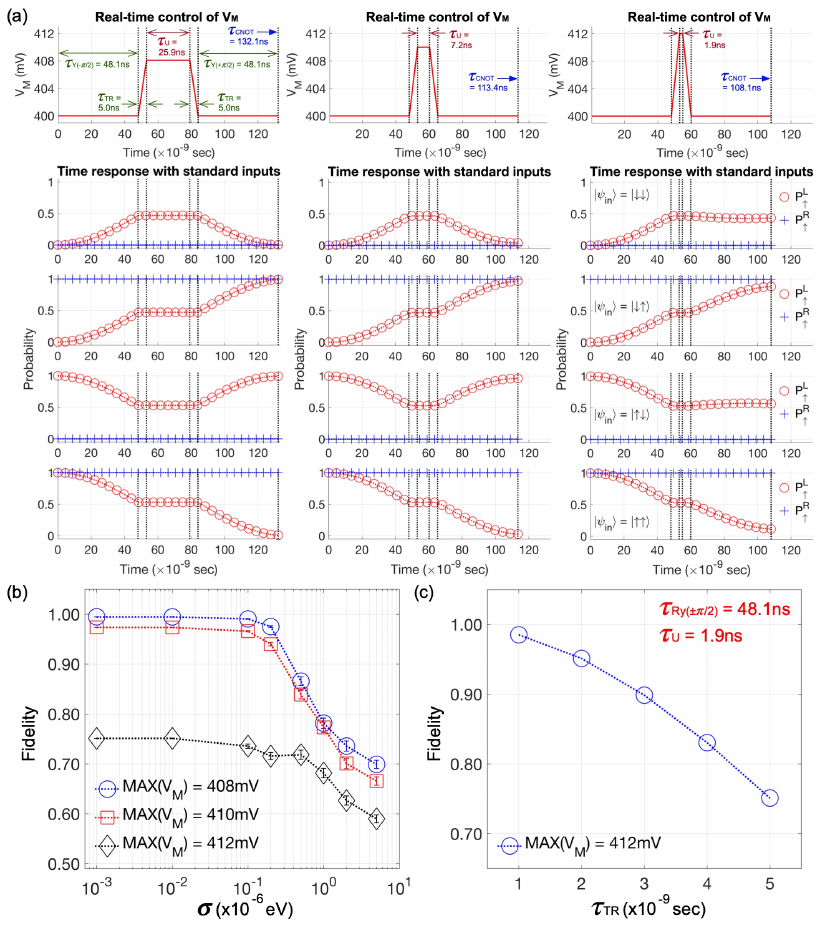}
\caption{\textbf{Acceleration of multi-step CNOT operation with $V_M$ control.} (a) 2-qubit time responses are simulated for the three cases, where the middle gate bias ($V_M$) is set to 408mV, 410mV, and 412mV to place the
	device in a regime of strong interaction. The reference case (408mV) takes 25.9 nanoseconds (nsec) to finish controlled-phase operation ($U$), and this time consumption becomes 7.2nsec and 1.9nsec when $V_M$ is 410mV
	and 412mV, respectively. (b) Corresponding fidelities are plotted as a function of the magnitude of charge noise, where we find the noise-driven degradation in fidelity of the second case (410mV) does not show remarkable
	difference compared to the reference case though 18.7 nsec can be saved for gating. The last case (412mV) is more robust to charge noises, but the average fidelity is not good even at $\sigma$ = 10$^{-3}$$\mu$eV (75.1\%)
	due to the transition of $V_M$ ($\tau_{TR}$ = 5nsec), as shown in (c) the loss in fidelity that is calculated as a function of $\tau_{TR}$.
	}
\label{fig:res4}	
\end{figure}

In Figure \ref{fig:res1}(b), we showed that the interaction energy between QDs has little effects on the resonance frequency of each spin qubit, so the gating time of $U$ can be safely controlled with $J$
(and thus $V_M$) with no worries for unintentional variations in any $E_{ZL}$- and $E_{ZR}$-related elements of the 4$\times$4 Heisenberg Hamiltonian \cite{R36}. With this background, we investigate
what happens on the noise-robustness of the multi-step CNOT operation if the gating time of $U$ is further reduced. For this purpose, the multi-step CNOT gate is simulated at $V_M$ = 410mV and 412mV,
where other control parameters are kept the same as the previously used ones. The time responses in Figure \ref{fig:res4}(a) clearly show that the entanglement is generated in 7.2nsec and 1.9nsec when
$V_M$ is 410mV and 412mV, respectively, and thus the CNOT gating time is reduced to 113.4nsec and 108.1nsec. Figure \ref{fig:res4}(b), which shows the fidelity of each noisy CNOT operation, indicates
that the noise-robustness at $V_M$ = 410mV does not quite change compared to the case of $V_M$ = 408mV though entanglement is generated must faster (25.9nsec $\rightarrow$ 7.2nsec). This result,
being different from the one obtained through a comparison between the single-step and the multi-step CNOT gate at $V_M$ = 408mV, can be explained with the fact that the time-integration of $J$ ($i.e.$,
\(\int_{0}^{\tau_U} J(t) \,dt\)) remains the same in the two cases (19.3MHz$\times$25.9nsec and 69.5MHz$\times$7.2nsec when $V_M$ is 408mV and 410mV, respectively), while, in the previous two cases
where $V_M$ is kept the same, the time-integration becomes smaller in the multi-step operation (19.3MHz$\times$25.9nsec) than in the single-step one (19.3MHz$\times$100.4nsec). If $V_M$ is increased
to 412mV, the time-integration still remains similar (266.1MHz$\times$1.9ns), showing $<$1\% deviation from the values at $V_M$ = 408mV and 410mV. In this case, however, the average fidelity gets worse
even under weak noises (75.1\% at $\sigma$ = 10$^{-3}$$\mu$eV), and this is due to the transition of $V_M$ that is essential to switch the interaction strength of QDs. Figure \ref{fig:res4}(c), which shows
the loss in fidelity of the multi-step CNOT operation at $V_M$ = 412mV as a function of $\tau_{TR}$, indicates that the loss can be reduced with a faster bias-transition, and we observe that the fidelity is
recovered back to 98.52\% if the transition can be conducted in 1nsec. Overall, it is fair to say that increasing the speed of U gating has little effects on the fidelity under charge noises, but still contributes
to saving the gating time, so, at $V_M$ = 410mV where the fidelity is not yet quite affected by a 5nsec-transition of $V_M$, the multi-step CNOT gate can be completed with just 10\% larger time-cost
(113.4nsec) than the single-step gate (100.4nsec).

\section{Conclusion}
Entangling logic operations under charge noises are computationally investigated in a silicon double quantum dot (DQD) system where quantum bits (qubits) are encoded to the confined electron spins.
Using a realistic DQD platform based on a silicon$/$silicon-germanium (Si$/$SiGe) heterostructure, we make a solid connection to the recent experimental work \cite{R10} where a fast controlled-X (CNOT)
gate has been implemented with a single-step control, but also extend the modeling scope into noise-driven behaviors of the single-step CNOT operation and 1-qubit rotations by incorporating random noisy
potential energies into device simulations. Though the 1-step implementation of a CNOT gate in the Si DQD platform has opened the fundamental pathway for securing a fast CNOT gate with simple controls,
it severely suffers from charge noises due to unintended fluctuations in the interaction energy between QDs, so its fidelity reaches lower than 35\% when the standard deviation of noisy potential energies
($\sigma$) is 5$\mu$eV. In contrast, 1-qubit rotations are generally quite robust to charge noises since the noisy fluctuation in potential distributions hardly affects the resonance frequency of individual spins.
Employing a DQD-native controlled-phase operation can be remarkably helpful for noise-robust implementation of a CNOT gate, because, at the same strength of 2-qubit interaction, it generates quantum
entanglement much faster than the single-step CNOT operation. Although additional 1-qubit rotations need to be conducted sequentially in time to complete the CNOT operation, they have little effects on
the noise-robustness, so the overall fidelity reaches $\sim$70\% at $\sigma$= 5$\mu$eV in spite of the increased complexity in device controls associated with additional 1-qubit rotations. Another benefit
of the controlled-phase operation implemented in the DQD platform is that its speed can be enhanced by increasing the strength of 2-qubit interaction with almost no degradation in noise-robustness. In
consequence, the associated CNOT gating can be conducted as fast as the single-step operation. Being supported with rigorous simulations, the engineering details discussed in this work can contribute
to elevating the current status of a Si QD platform for robust designs of scalable quantum processors.

\section*{Acknowledgements}
This work has been supported by the Korea Institute of Science and Technology Information (KISTI) institutional R\&D program (K-22-L02-C09) and by the grant from the Institute for Information \& Communications
Technology Promotion (2019-0-00003) funded by the Korea government (MSIP). The NURION high performance computing resource has been extensively utilized for simulations.

\begin{thebibliography}{10}
\expandafter\ifx\csname url\endcsname\relax
  \def\url#1{\texttt{#1}}\fi
\expandafter\ifx\csname urlprefix\endcsname\relax\def\urlprefix{URL }\fi
\expandafter\ifx\csname href\endcsname\relax
  \def\href#1#2{#2} \def\path#1{#1}\fi

\bibitem{R01}
T.~Kobayashi, J.~Salfi, C.~Chua, J.~van~der Heijden, M.~House, D.~Culcer,
  W.~Hutchison, B.~Johnson, J.~McCallum, H.~Riemann, N.~Abrosimov, P.~Becker,
  H.~Pohl, M.~Y. Simmons, S.~Rogge, {Engineering long spin coherence times of
  spin-orbit qubits in silicon}, Nature Materials 20 (2021) 38--42.

\bibitem{R02}
J.~T. Muhonen, J.~P. Dehollain, A.~Laucht, F.~E. Hudson, R.~Kalra,
  T.~Sekiguchi, K.~M. Itoh, D.~N. Jamieson, J.~C. McCallum, A.~S. Dzurak,
  A.~Morello, {Storing quantum information for 30 seconds in a nanoelectronic
  device}, Nature Nanotechnology 9 (2014) 986--991.

\bibitem{R03}
M.~Veldhorst, J.~C.~C. Hwang, C.~H. Yang, A.~W. Leenstra, B.~de~Ronde, J.~P.
  Dehollain, J.~T. Muhonen, F.~E. Hudson, K.~M. Itoh, A.~Morello, A.~S. Dzurak,
  {An addressable quantum dot qubit with fault-tolerant control-fidelity},
  Nature Nanotechnology 9 (2014) 981--985.

\bibitem{R04}
E.~Kawakami, T.~Jullien, P.~Scarlino, D.~R. Ward, D.~E. Savage, M.~G. Lagally,
  V.~V. Dobrovitski, M.~Friesen, S.~N. Coppersmith, M.~A. Eriksson, L.~M.~K.
  Vandersypen, {Gate fidelity and coherence of an electron spin in an Si$/$SiGe
  quantum dot with micromagnet}, Proceedings of the National Academy of
  Sciences of the United States of America 113 (2016) 11738--11743.

\bibitem{R05}
E.~Kawakami, P.~Scarlino, D.~R. Ward, F.~R. Braakman, D.~E. Savage, M.~G.
  Lagally, M.~Friesen, S.~N. Coppersmith, M.~A. Eriksson, L.~M.~K. Vandersypen,
  {Electrical control of a long-lived spin qubit in a Si$/$SiGe quantum dot},
  Nature Nanotechnology 9 (2014) 666--670.

\bibitem{R06}
K.~Takeda, J.~Kamioka, T.~Otsuka, J.~Yoneda, T.~Nakajima, M.~R. Delbecq,
  S.~Amaha, G.~Allison, T.~Kodera, S.~Oda, S.~Tarucha, {A fault-tolerant
  addressable spin qubit in a natural silicon quantum dot}, Science Advances 2
  (2016) e1600694.

\bibitem{R07}
J.~Yoneda, K.~Takeda, T.~Otsuka, T.~Nakajima, M.~R. Delbecq, G.~Allison,
  T.~Honda, T.~Kodera, S.~Oda, Y.~Hoshi, N.~Usami, K.~M. Itoh, S.~Tarucha, {A
  quantum-dot spin qubit with coherence limited by charge noise and fidelity
  higher than 99.9\%}, Nature Nanotechnology 13 (2018) 102--106.

\bibitem{R08}
A.~J. Sigillito, M.~J. Gullans, L.~F. Edge, M.~Borselli, J.~R. Petta, {Coherent
  transfer of quantum information in a silicon double quantum dot using
  resonant SWAP gates}, npj Quantum Information 5 (2019) 110.

\bibitem{R09}
T.~F. Watson, S.~G.~J. Philips, E.~Kawakami, D.~R. Ward, P.~Scarlino,
  M.~Veldhorst, D.~E. Savage, M.~G. Lagally, M.~Friesen, S.~N. Coppersmith,
  M.~A. Eriksson, L.~M.~K. Vandersypen, {A programmable two-qubit quantum
  processor in silicon}, Nature 555 (2018) 633--637.

\bibitem{R10}
D.~M. Zajac, A.~J. Sigillito, M.~Russ, F.~Borjans, J.~M. Taylor, G.~Burkard,
  J.~R. Petta, {Resonantly driven CNOT gate for electron spins}, Science 359
  (2018) 439--442.

\bibitem{R11}
W.~Huang, C.~H. Yang, K.~W. Chan, T.~Tanttu, B.~Hensen, R.~C.~C. Leon, M.~A.
  Fogarty, J.~C.~C. Hwang, F.~E. Hudson, K.~M. Itoh, A.~Morello, A.~Laucht,
  A.~S. Dzurak, {Fidelity benchmarks for two-qubit gates in silicon}, Nature
  569 (2019) 532--536.

\bibitem{R12}
W.~K. Wootters, W.~S. Leng, {Quantum Entanglement as a Quantifiable Resource},
  Philosophical Transactions of the Royal Society A 356 (1998) 1717--1731.

\bibitem{R13}
R.~Horodecki, P.~Horodecki, M.~Horodecki, K.~Horodecki, {Quantum entanglement},
  Reviews of Modern Physics 81 (2009) 865--942.

\bibitem{R14}
C.~H. Bennett, G.~Brassard, C.~Cr$\acute{e}$peau, R.~Jozsa, A.~Peres, W.~K.
  Wootters, {Teleporting an unknown quantum state via dual classical and
  einstein-podolskyrosen channels}, Physical Review Letters 70 (1993) 1895.

\bibitem{R15}
P.~W. Shor, {Algorithms for quantum computation: discrete logarithms and
  factoring}, in: Proceedings of the Annual Symposium on Foundations of
  Computer Science, 1994, pp. 124--134.
\newblock \href {http://dx.doi.org/10.1109/SFCS.1994.365700}
  {\path{doi:10.1109/SFCS.1994.365700}}.

\bibitem{R16}
D.~Lachance-Quirion, S.~Wolski, Y.~Tabuchi, S.~Kono, K.~Usami, Y.~Nakamura,
  {Entanglement-based single-shot detection of a single magnon with a
  superconducting qubit}, Science 367 (2020) 425--428.

\bibitem{R17}
L.~M.~K. Vandersypen, I.~L. Chuang, {NMR techniques for quantum control and
  computation}, Reviews of Modern Physics 76 (2005) 1037--1069.

\bibitem{R18}
A.~V. Kuhlmann, J.~Houel, A.~Ludwig, L.~Greuter, D.~Reuter, A.~D. Wieck,
  M.~Poggio, R.~J. Warburton, {Charge noise and spin noise in a semiconductor
  quantum device}, Nature Physics 9 (2013) 570--575.

\bibitem{R19}
E.~J. Connors, J.~Nelson, H.~Qiao, L.~F. Edge, J.~M. Nichol, {Low-frequency
  charge noise in Si$/$SiGe quantum dots}, Physical Review B 100 (2019) 165305.

\bibitem{R20}
C.~D. Wilen, S.~Abdullah, N.~Kurinsky, C.~Stanford, L.~Cardani,
  G.~$\acute{D}$Imperio, C.~Tomei, L.~Faoro, L.~Loffe, C.~Liu, A.~Opremcak,
  B.~Christensen, J.~DuBois, R.~McDermott, {Correlated charge noise and
  relaxation errors in superconducting qubits}, Nature 594 (2021) 369--373.

\bibitem{R21}
S.~Pezzagna, J.~Meijera, {Quantum computer based on color centers in diamond},
  Applied Physics Reviews 8 (2021) 011308.

\bibitem{R22}
L.~Kranz, S.~K. Gorman, B.~Thorgrimsson, Y.~He, D.~Keith, J.~G. Keizer, M.~Y.
  Simmons, {Exploiting a Single-Crystal Environment to Minimize the Charge
  Noise on Qubits in Silicon}, Advanced Materials 32 (2020) 2070298.

\bibitem{R23}
R.~Maurand, X.~Jehl, D.~Kotekar-Patil, A.~Corna, H.~Bohuslavskyi,
  R.~Lavi\'{e}ville, L.~Hutin, S.~Barraud, M.~Vinet, M.~Sanquer, S.~Franceschi,
  {A CMOS silicon spin qubit}, Nature Communications 7 (2016) 13575.

\bibitem{R24}
V.~Mazzocchi, P.~Sennikov, A.~Bulanov, M.~Churbanov, B.~Bertrand, L.~Hutin,
  J.~Barnes, M.~Drozdov, J.~Hartmann, M.~Sanquer, {99.992\% $^{28}$Si CVD-grown
  epilayer on 300mm substrates for large scale integration of silicon spin
  qubits}, Journal of Crystal Growth 509 (2019) 1--7.

\bibitem{R25}
M.~Reed, B.~Maune, R.~Andrews, M.~Borselli, K.~Eng, M.~Jura, A.~Kiselev,
  T.~Ladd, S.~Merkel, I.~Milosavljevic, E.~Pritchett, M.~Rakher, R.~Ross,
  A.~Schmitz, A.~Smith, J.~Wright, M.~Gyure, A.~Hunter, {Reduced Sensitivity to
  Charge Noise in Semiconductor Spin Qubits via Symmetric Operation}, Physical
  Review Letters 116 (2016) 110402.

\bibitem{R26}
J.~Kang, J.~Ryu, H.~Ryu, Exploring the behaviors of electrode-driven si quantum
  dot systems: from charge control to qubit operations, Nanoscale 13 (2021)
  332--339.

\bibitem{R27}
Y.~Nosho, Y.~Ohno, S.~Kishimoto, T.~Mizutani, {Relation between conduction
  property and work function of contact metal in carbon nanotube field-effect
  transistors}, Nanotechnology 17 (2006) 3412--3415.

\bibitem{R28}
J.~Wang, A.~Rahman, A.~Ghosh, G.~Klimeck, M.~Lundstrom, On the validity of the
  parabolic effective-mass approximation for the {I-V} calculation of silicon
  nanowire transistors, IEEE Transactions on Electron Devices 52 (2005)
  1589--1595.

\bibitem{R29}
R.~Neumann, L.~R. Schreiber, {Simulation of micro-magnet stray-field dynamics
  for spin qubit manipulation}, Journal of Applied Physics 117 (2015) 193903.

\bibitem{R30}
J.~Yoneda, T.~Otsuka, T.~Nakajima, T.~Takakura, T.~Obata,
  M.~Pioro-Ladri$\grave{e}$re, H.~Lu, C.~J. Palmstr$\o$m, A.~C. Gossard,
  S.~Tarucha, {Fast Electrical Control of Single Electron Spins in Quantum Dots
  with Vanishing Influence from Nuclear Spins}, Physical Review Letters 113
  (2014) 267601.

\bibitem{R31}
J.~Yoneda, T.~Otsuka, T.~Takakura, M.~Pioro-Ladri$\grave{e}$re, R.~Brunner,
  H.~Lu, T.~Nakajima, T.~Obata, A.~Noiri, C.~J. Palmstr$\o$m, A.~C. Gossard,
  S.~Tarucha, {Robust micromagnet design for fast electrical manipulations of
  single spins in quantum dots}, Applied Physics Express 8 (2015) 084401.

\bibitem{R32}
B.~Thorgrimsson, D.~Kim, Y.-C. Yang, L.~W. Smith, C.~B. Simmons, D.~R. Ward,
  R.~H. Foote, J.~Corrigan, D.~E. Savage, M.~G. Lagally, M.~Friesen, S.~N.
  Coppersmith, M.~A. Eriksson, {Extending the coherence of a quantum dot hybrid
  qubit}, npj Quantum Information 3 (2017) 32.

\bibitem{R33}
M.~A. Fogarty, K.~W. Chan, B.~Hensen, W.~Huang, T.~Tanttu, C.~H. Yang,
  A.~Laucht, M.~Veldhorst, F.~E. Hudson, K.~M. Itoh, D.~Culcer, T.~D. Ladd,
  A.~Morello, A.~S. Dzurak, {Integrated silicon qubit platform with single-spin
  addressability, exchange control and single-shot singlet-triplet readout},
  Nature Communications 9 (2018) 4370.

\bibitem{R34}
Z.~Shi, C.~B. Simmons, D.~R. Ward, J.~R. Prance, R.~T. Mohr, T.~S. Koh, J.~K.
  Gamble, X.~Wu, D.~E. Savage, M.~G. Lagally, S.~N.~C. M.~Friesen, M.~A.
  Eriksson, {Coherent quantum oscillations and echo measurements of a Si charge
  qubit}, Physical Review B 88 (2013) 075416.

\bibitem{R35}
X.~Wu, D.~R. Ward, J.~R. Prance, D.~Kim, J.~K. Gamble, R.~T. Mohr, Z.~Shi,
  D.~E. Savage, M.~G. Lagally, M.~Friesen, S.~N. Coppersmith, M.~A. Eriksson,
  {Two-axis control of a singlet-triplet qubit with an integrated micromagnet},
  Proceedings of the National Academy of Sciences of the United States of
  America 111 (2014) 11938--11942.

\bibitem{R36}
M.~Russ, D.~M. Zajac, A.~J. Sigillito, F.~Borjans, J.~M. Taylor, J.~R. Petta,
  G.~Burkard, {High-fidelity quantum gates in Si/SiGe double quantum dots},
  Physical Review B 97 (2018) 085421.

\end{thebibliography}
\end{document}